\documentclass[lettersize,journal]{IEEEtran}
\usepackage{amsmath,amsfonts}
\usepackage{algorithmic}
\usepackage{algorithm}
\usepackage{array}
\usepackage[caption=false,font=normalsize,labelfont=sf,textfont=sf]{subfig}
\usepackage{textcomp}
\usepackage{balance}
\usepackage{stfloats}
\usepackage{url}
\usepackage{verbatim}
\usepackage{graphicx}
\usepackage{xcolor}
\usepackage{cite}
\usepackage{multirow}
\begin{document}

\title{Wireless Powering Efficiency \\ of Deep-Body Implantable Devices}

\author{Icaro V. Soares, \IEEEmembership{Student Member, IEEE}, Mingxiang Gao, Erdem Cil, \IEEEmembership{Student Member, IEEE},\\ Zvonimir Šipuš, \IEEEmembership{Senior Member, IEEE},  Anja K. Skrivervik, John S. Ho, \IEEEmembership{Member, IEEE}, and~Denys~Nikolayev,~\IEEEmembership{Member, IEEE}
\thanks{Manuscript received October 7, 2022, revised November 25, 2022. This work was supported in part by the French \textit{Agence Nationale de la Recherche} (ANR) under grant ANR-21-CE19-0045 (project ``MedWave''), in part by the French Region of Brittany through the \textit{Stratégie d’attractivité durable} (SAD) project ''EM-NEURO.`` (\textit{Corresponding author: D.~Nikolayev}, denys.nikolayev@deniq.com)}%
\thanks{I. V. Soares and D. Nikolayev are with the Univ Rennes, CNRS, IETR – UMR 6164, FR-35000 Rennes, France.}
\thanks{M. Gao and A. K. Skrivervik are with the Microwave and Antenna Group, Ecole Polytechnique Fédérale de Lausanne, CH-1015 Lausanne, Switzerland.}
\thanks{E. Cil is with the Univ Rennes, CNRS, IETR~-- UMR 6164, FR-35000 Rennes, France and with BodyCAP, FR-14200 Hérouville St Clair, France.}
\thanks{Z. Šipuš is with the Faculty of Electrical Engineering and Computing, University of Zagreb, 10000 Zagreb, Croatia.}
\thanks{J. S. Ho is with the Department of Electrical and Computer Engineering, Institute for Health Innovation \& Technology, National University of Singapore, Singapore 117599, and also with the N.1 Institute for Health,
National University of Singapore, Singapore 117599.}
\thanks{Color versions of one or more of the figures in this paper are available online at http://ieeexplore.ieee.org.}}

\markboth{IEEE TRANSACTIONS ON MICROWAVE THEORY AND TECHNIQUES}%
{Soares \MakeLowercase{\textit{et al.}}: Wireless Powering Efficiency of Deep-Body Implantable Devices}


\maketitle

\begin{abstract}
\textcolor{black}{The wireless power transfer efficiency to implanted bioelectronic devices is constrained by several frequency-dependent physical mechanisms. Recent works have developed several mathematical formulations to understand these mechanisms and predict the optimal operating conditions. However, existing approaches rely on simplified body models, which are unable to capture important aspects of wireless power transfer. Therefore, this paper proposes the efficiency analysis approach in anatomical models that can provide insightful information on achieving the optimum operation conditions. First, this approach is validated with a theoretical spherical wave expansion analysis, and the results for a simplified spherical model and a human pectoral model are compared. The results show that although a magnetic receiver outperforms an electric one for near-field operation and both sources could be equally employed in far-field range, it is in mid-field that the maximum efficiency is achieved with an optimum frequency between 1--5 GHz depending on the implantation depth. The receiver orientation is another factor that affects the efficiency, with a maximum difference between the best and worst-case scenarios around five times for the electric source and over 13 times for the magnetic one. This approach is used to analyze the case of a deep-implanted pacemaker wirelessly powered by an on-body transmitter and subjected to stochastic misalignments. We evaluate the efficiency and exposure, and we demonstrate how a buffered transmitter can be tailored to achieve maximum powering efficiency. Finally, design guidelines that lead to optimal implantable wireless power transfer systems are established from the results obtained with the proposed approach.}
\end{abstract}

\begin{IEEEkeywords}
Biomedical electronics, implantable devices, pacemakers, radiation efficiency, wireless power transfer.
\end{IEEEkeywords}

\vspace{2cm}

\section{Introduction}\label{secI}
\IEEEPARstart{W}{ireless} Power Transfer (WPT) has been the subject of many works and publications mainly due to its technological and commercial appeal. It was founded in the works of Nikola Tesla, who  first proposed the use of electromagnetic waves to wirelessly power electric devices. Even though the theoretical aspects have been studied throughout the years and are well consolidated, the physical realization and the proposition of different techniques for its application have seen renewed interest with the rise of portable electronic devices~\cite{1}. 

One way to classify the WPT techniques that use electromagnetic waves as the energy carrier is by the radiation region that comprises the distance between the transmitter and receiver. For instance, the near-field techniques are based on the near-field coupling between antennas that can be either inductive or capacitive \cite{2}.

Consequently, due to the short distance between antennas, these techniques can achieve high efficiency, and they are able to deliver high power levels \cite{3}. This efficiency can be even further increased by electromagnetic resonance \cite{4}. On the other side, the radiative or far-field techniques can cover a wide distance range with the drawback of lower efficiency. Therefore, they usually employ highly directive antennas to reduce the power dispersed in free-space and thus, increase the received power \cite{5}. A new mid-field WPT technique has been proposed to achieve higher efficiency than the far-field techniques and larger distance ranges than their near-field counterparts. In mid-field WPT, the distance between transmitter and receiver is in the order of a wavelength, and, in this case, the electromagnetic fields are composed of reactive and propagating components \cite{6},\cite{7}.

In the context of biomedical engineering, the application of WPT for implantable bioelectronic and biosensor devices is envisaged to monitor physiological processes \cite{8}, drug delivery \cite{9}, and organ stimulation through electric signals \cite{10}, to name a few. Therefore, different WPT techniques have already been applied for wireless powering bioelectronic implants~\cite{barbruni_miniaturised_2020}. For instance, near-field WPT is employed for subcutaneous pacemakers \cite{11,12,49}, neural and spinal cord stimulators \cite{13,47,48}, cochlear \cite{14}, and ocular implants \cite{15}, among other. All these applications share the same characteristic of having a small implantation depth. Conversely, far-field techniques are used for ultra-low-power implants usually applied for biotelemetry \cite{8}. However, for powering deep-implanted devices, the mid-field WPT is mostly used. For example, it has been applied for cardiac implants and pacemakers \cite{16,17,18}, gastrointestinal endoscopy capsules \cite{19}, \cite{20}, and neural implants \cite{21}.

Several physical, technical, and legal aspects impose limitations on the in-body WPT application. First, the body tissues are composed of dynamic, highly heterogeneous, dispersive, and lossy media that deteriorate the efficiency of WPT techniques which in other applications present a reasonable efficiency. The two prominent factors that reduce WPT efficiency are the electromagnetic wave attenuation in these lossy tissues, which increases with the operating frequency, and the reflection at the interface between different media \cite{22}. Moreover, the miniaturized implantable receivers are usually both physically and electrically small, further reducing the maximum achievable efficiencies \cite{23}, not to mention other sources of losses such as impedance mismatching, misalignments, and internal resonances in the body organs. Another major concern is the human body exposure to electromagnetic fields, which must comply with regulatory limits for human use~\cite{24}. Therefore, the WPT system parameters -– such as operating frequency, nature of the source, and implantation depth -– must be judiciously chosen in order to achieve the highest efficiency possible, taking into account the exposure level regulations.

{\color{black} Several models have been proposed to mathematically assess the maximum radiation efficiency and optimal frequency for implantable bioelectronic devices. Simplified analytical models can help making conclusions about the optimal parameters of the system independently from the source and receiver design~\cite{berkelmann_antenna_2022} -- these conclusions are difficult to make with 3D full-wave simulations because it cannot search through the entire design space. The planar stratified heterogeneous model \cite{25},\cite{26} is a simple approach that allows an analytical formulation that includes the wave reflection between different layers.  In this model, the equivalence principle replaces a volumetric surface current distribution for an in-plane electric current density source that produces the same electromagnetic fields. Although this model is mathematically more straightforward and provides physical insight of the problem under analysis, it considers the human body as planar and infinite media. Therefore, it disregards the effects of the human body's shape and dimensions on radiation performance. Besides, the assumption of an infinite medium may lead to inaccuracies in the maximum efficiency prediction.}

{\color{black} On the other hand, the spherical wave expansion \cite{27,28,29} is a semi-analytical approach that represents the human body as a spherical model composed of a homogeneous or concentric multilayered medium. In this case, the source is represented by an electric or magnetic dipole encapsulated by a small sphere positioned anywhere inside the spherical phantom. Then, the electric and magnetic fields are decomposed in vector spherical harmonics, and, from orthogonality relationships, the problem is mathematically solved by a linear system. From the computational point of view, this approach is less costly than a full-wave analysis and can lead to insightful results once the implantation depth mainly influences the level of achievable power density just outside the body. Furthermore, due to the high electric permittivity of the biological tissues, the near-field region is shorter than in free space; therefore, the radiation pattern of mm-sized implantable antennas is less sensitive to the shape of the phantom itself \cite{29}. However, this approach is difficult to implement for more complex sources and realistic body models. Apart from that, the standing wave pattern in spherical models differs from the anatomical-shaped ones. Therefore, this difference is also verified in the WPT efficiency values obtained with both phantoms. For this intent, a full-wave computation needs to be carried out to solve the inhomogeneous wave equation numerically given an arbitrary finite-sized source \cite{23}.}

This work aims to investigate the electromagnetic energy exchange between an optimal on-body transmitter and an implantable receiver, as well as to analyze the influence of the system's parameters on the WPT efficiency. We develop a general 2D-axisymmetric full-wave approach is, providing the best-case scenario for the energy transfer efficiency as a function of the frequency, the implantation depth, and the nature of the receiver, i.e., electric or magnetic. The basis for this problem formulation and some results for a simplified spherical phantom model have been presented in \cite{30}. However, the proposed approach is further extended and validated with a theoretical model based on spherical wave expansion in this paper. Then, both methods are used to evaluate the frequency-dependent radiation efficiency for several implantation depths considering electric and magnetic receivers, and these results surpass the efficiency levels previously presented in the literature with other approaches.

Previous studies on modeling electromagnetic wave propagation in the human body mainly focused on wireless communications with implantable antennas \cite{28},\cite{31} medical treatments based on electromagnetic waves such as hyperthermia \cite{32,33,34,35} or dosimetry \cite{36}. However, the investigation of wave propagation for WPT applications \cite{16},\cite{25} involves a specific focus on the electromagnetic interaction between an on-body transmitter and an in-body receiver. Therefore, the optimal operating conditions identified for in-body biotelemetry and hyperthermia applications might differ. Specifically, for the in-body biotelemetry, the goal is to maximize the omnidirectional radiation performance through the high-contrast tissue–air interface \cite{22}. For hyperthermia, the main objective is to control the local and non-linear [i.e., $\hat{\varepsilon}(\textbf{E})$)] absorption of energy close to the source \cite{33}. As a result, most deep-body implantable devices operate in the MedRadio band  (401–457 MHz) \cite{37}; the optimal band for hyperthermia is between 130 MHz and 500 MHz \cite{38}, whereas for WPT, the maximum efficiency is achieved around 1–3 GHz \cite{25},\cite{26},\cite{30},\cite{30_1}. This work studies the electromagnetic exchange in an in-/on-body WPT system, derives, and demonstrates the optimal operating conditions in terms of efficiency and exposure as a function of implantation depth and nature of the power source antenna. 

{\color{black} Moreover, this study applies for the first time the 2D-axisymmetric full-wave formulation to assess the WPT efficiency in an anatomical scenario considering a deep-implanted mm-sized pacemaker in a transverse cut model for the human pectoral. The proposed approach has the advantage of taking into account the complexities of the human body regarding its shape, dimensions, and electromagnetic heterogeneity, being less computationally costly than a tridimensional full-wave analysis. In this way, it is possible to analyze the real-life application of WPT for powering a cardiac stimulator as presented in \cite{16}. In addition, different configurations for electric and magnetic on-body sources are investigated, and the results for the anatomical model are compared with the simplified spherical phantom. Finally, the origin for the losses is discussed from these results, and ways to mitigate them and achieve maximum efficiency are investigated from parametric analysis of on-body transmitter that wirelessly powers the implanted pacemaker.}

\section{Efficiency Analysis with Simplified Spherical~Models}\label{secII}

\begin{figure*}[b]
\centering  
\includegraphics[width=\textwidth]{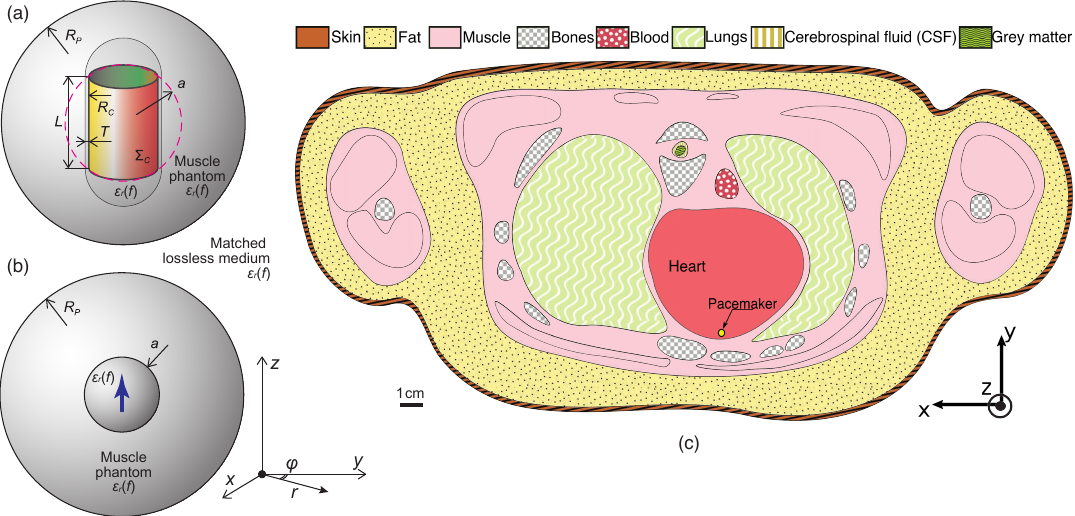}
\caption{Geometrical representation of the problem under analysis: (a-b) simplified spherical model (not to scale). A muscle-equivalent spherical phantom with radius $R_p$ and dispersive complex permittivity $\hat{\varepsilon}_r(f)$, surrounded by a lossless matched medium with ${\varepsilon}_r(f)$, encloses the implant. Both (a) and (b) share the same axis and are surrounded by the same matched medium. For the: (a) 2D-axisymmetric model, this implant is insulated by a capsule-shaped volume with ${\varepsilon}_r(f)$ and the receiving electric or magnetic antenna is modeled as their corresponding current distribution $\textbf{J}_s$ on the cylindrical aperture, whereas for the (b) spherical wave expansion model, the source is a lossless bubble with ${\varepsilon}_r(f)$ and radius a, and the receiver is an elementary electric or magnetic dipole. (c) Anatomical 2D model of a human pectoral region comprising nine dispersive tissues. An on-body source represents the transmitter, and the receiver is a deep-tissue pacemaker implanted inside the heart at a depth of 37 mm from the closest skin interface and filled by a lossless dielectric with ${\varepsilon}_r(f)$, matched to the wave-impedance in the myocardium.}
\label{fig1}
\end{figure*}

{\color{black} The problem under analysis consists of a WPT system composed of an implantable bioelectronic device that acts as a receiver and a transmitter structure conformal to the surface of the body. To determine the achievable WPT efficiency, the human body is modeled as a dispersive homogeneous spherical phantom of radius $R_p$ with complex permittivity $\hat{\varepsilon}_r(f)$ equivalent to the human muscle tissue \cite{39}, as shown in Fig.~\ref{fig1}(a). The phantom radius $R_p$ is directly related to the implantation depth, and its evaluated range embraces most deep-body implantable WPT applications. Inside this phantom, the implantable device is represented as a lossless $\varepsilon_r(f)$ capsule-shaped volume that is matched to the wave impedance in the surrounding tissue and has length $L$ and radius $R_c$. This spherical model allows the results obtained with the proposed approach be compared with semi-analytical solutions based on spherical wave expansion. In addition, this simplified model disregards the shape complexities and heterogeneities leading to more general results that are better suitable for an initial analysis of implantable WPT systems.}

To eliminate losses due to reflection at the interface between the body and the external medium and thereby ensure the maximum WPT efficiency, the permittivity of the region around the phantom is considered as the real part of  $\hat{{\varepsilon}}_r(f)$, i.e., the external medium is lossless ${{\varepsilon}}_r(f)$ and matched to the wave impedance in the muscle. In this way, the on-body transmitter is well matched to the tissue, which can be physically realized using a high dielectric constant layer. This model disregards any impedance mismatching in the antennas and any system-level losses in electronic circuitry.

To mathematically formulate the problem, an equivalent problem can be proposed based on reciprocity \cite{40}. For this intent, the radiation source is modeled as the electric current density distribution $\textbf{J}_s$ inside the phantom, as indicated in Fig. \ref{fig1}(a). The reciprocity principle can be applied since the medium is assumed linear. Consequently, the radiated fields induce a power flow in the lossy phantom and, from energy conservation, the transmitted power $P_t$ is given by:
\begin{equation}\label{eq1}
  P_t=P_r+P_d+i2\omega(\overline{W_m}-\overline{W_e})
\end{equation}
where $P_r$ and $P_d$ are respectively the received and dissipated powers, $\overline{W_m}$ and $\overline{W_e}$ are the time-average magnetic and electric energies, respectively.

From the time-harmonic electric $\textbf{E}$ and magnetic $\textbf{H}$ fields, as well as the source electric $\textbf{J}_s$ and magnetic $\textbf{M}_s$ current densities, each one of the total power components can be calculated as follow:
\begin{subequations}
\begin{equation}\label{eq2a}
  P_t = \oint_{\Omega_s}\left ( \frac{1}{2}\textbf{E}\times\textbf{H}^* \right ) \cdot \textup{d}\mathbf{s}
\end{equation}    
\begin{equation}\label{eq2b}
  P_r = \oint_{\Omega_p}\left ( \frac{1}{2}\textbf{E}\times\textbf{H}^* \right ) \cdot \textup{d}\mathbf{s}
\end{equation}
\begin{equation}\label{eq2c}
  P_d = \frac{1}{2}\int_{\Omega_p}\sigma\left | \textbf{E} \right |^2\;\textup{d}v
\end{equation}
\end{subequations}
where $\sigma$ is the phantom's electric conductivity which is related to the imaginary part of the phantom’s complex permittivity $\Im\left [ \hat{\varepsilon}_r(f) \right ]$ through $\sigma\left(f\right)=2\pi\varepsilon_0 f \Im\left[ \hat{\varepsilon}_r\left(f\right) \right]$. The integration domains $\Omega_s$ and $\Omega_p$ indicate that the integral is performed on the surface of the source and the phantom, respectively. Besides, the integrals (\ref{eq2a}-\ref{eq2c}) can be considerably simplified due to the problem's axial symmetry. Finally, the WPT efficiency $\eta$ is defined as:
\begin{equation}\label{eq3}
  \eta \equiv \frac{\Re(P_r)}{\Re(P_t)} = 1-\frac{\Re(P_d)}{\Re(P_t)} 
\end{equation} 
with $\Re$ being the real part of the complex variable inside the brackets. 

Hereafter, the implantable bioelectronic device is referred to as the receiver antenna, whereas the transmitter antenna is the radiating on-body structure.

\subsection{Full-wave 2D-Axisymmetric Analysis}

{\color{black} In order to numerically evaluate the WPT efficiency using the proposed approach, this model was implemented using the full-wave 2D-axisymmetric formulation available in COMSOL Multiphysics (assuming no variation of the fields in the $\varphi$ direction). Electric and magnetic sources with size $L = 1\;\textup{cm}$ and $R_c = L/3$ are considered in this computation, which corresponds to the standard encapsulation size used for implantable and ingestible bioelectronic devices \cite{37}. Each source is represented by the electric $\textbf{J}_s^E(r,\varphi,z)$ and magnetic $\textbf{J}_s^H(r,\varphi,z)$ surface current densities given by \cite{23}:}
\begin{subequations}
\begin{equation}\label{eq4a}
  \textbf{J}_s^E = \left [ 0,0,\cos\left (\frac{\pi z}{L}  \right ) \right ]
\end{equation}    
\begin{equation}\label{eq4b}
  \textbf{J}_s^H = \left [  0,1,0  \right ].
\end{equation}
\end{subequations}
Then, the WPT efficiency in (\ref{eq3}) can be calculated from the evaluated integrals in (\ref{eq2a}-\ref{eq2c}).

\subsection{Spherical Wave Expansion Analysis}

The results from a semi-analytical approach based on spherical wave modal expansion are used as a reference to validate the proposed model. In this model depicted in Fig. \ref{fig1}(b), the electromagnetic fields \textbf{E} and \textbf{H} in a spherical structure free of charges can be expressed using vector spherical harmonics \textbf{M} and \textbf{N} \cite{40}:
\begin{equation}\label{eq5}
  \begin{Bmatrix} \textbf{E}\\ -i\eta \textbf{H} \end{Bmatrix} = \sum_{n,m}a_{mn}\begin{Bmatrix} \textbf{M}_{mn}\\ \textbf{N}_{mn} \end{Bmatrix} +b_{mn}\begin{Bmatrix} \textbf{N}_{mn}\\ \textbf{M}_{mn} \end{Bmatrix}
\end{equation} 
with
\begin{subequations}
\begin{equation}\label{eq6a}
  \textbf{M}_{mn}=\mathbf{\nabla}\times \textbf{r}\psi_{mn}
\end{equation}    
\begin{equation}\label{eq6b}
  \textbf{N}_{mn}=\frac{1}{k}\mathbf{\nabla}\times\mathbf{\nabla}\times \textbf{r}\psi_{mn}
\end{equation}
\end{subequations}
where $k$ denotes the wave number of the considered media, $\textbf{r}$ is the unit position vector in the $r$-direction, and $\psi_{mn}$ is the solution to the Helmholtz differential equation:
\begin{equation}\label{eq7}
  \psi_{mn}=\frac{1}{kr} Z_{n}(kr)P_{n}^{m}(\cos\theta)e^{im\varphi}. 
\end{equation}

The Schelkunoff type spherical Bessel or Hankel functions are denoted by $Z_n$, whereas $P_n^m$ are the associated Legendre functions of the first kind \cite{41}. From the orthogonality relationships for spherical harmonics, (\ref{eq5}) becomes a linear system whose solution provides the modal representation for the electromagnetic fields. More details about this implementation can be found in \cite{28}.

The integrals in (\ref{eq2a}-\ref{eq2c}) can be evaluated from the spherical model coefficients and vectors and, thus, the efficiency can be obtained through (\ref{eq3}). This theoretical approach can also be applied to off-centered sources \cite{27},\cite{28}. 

\subsection{Experimental Validation}\label{sec_ExpVal}

\begin{figure}[htbp!]
\centering  
\includegraphics[scale = 1]{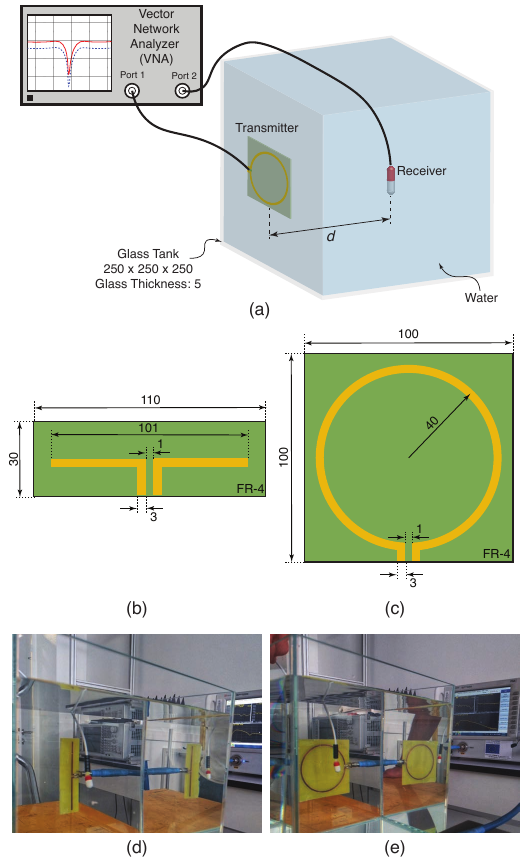}
\caption{\textcolor{black}{Experimental evaluation of WPT efficiency in lossy media: (a) the transmitter is positioned outside a cubic glass tank filled with water, whereas the receiver designed in \cite{30_1} is inside the tank at a distance $d$ from the transmitter. Then, the scattering parameters are measured with a two-port VNA considering (b--d) an electric (dipole antenna), and (c--e) a magnetic (loop antenna) transmitter. All dimensions are in millimeters.}}
\label{fig1b}
\end{figure}

\textcolor{black}{In order to experimentally evaluate the WPT efficiency of devices deeply-implanted in lossy media, the measurement setup described in Fig. \ref{fig1b} is proposed. It consists of a transmitter positioned on the outside wall of a cubic glass tank filled with water. The choice of water as the medium for the measurements is based on the fact that its dielectric properties are well-characterized across all analyzed frequency band; therefore, it is best suitable for validating the proposed approach. In contrast, the receiver is placed inside the tank at a distance $d$ from the transmitter, with aligned polarization. For these measurements, two kinds of transmitters were used: a dipole and a loop antenna with the dimensions shown in Fig. \ref{fig1b}(b) and Fig. \ref{fig1b}(c), respectively. On the other side, the receiver design is described in \cite{30_1}. The results obtained with this experiment setup are expected to be representative of wirelessly-powered devices implanted in biological media primarily composed of soft human tissues. At the same time, the glass permittivity can numerically approximate those of hard tissues.} 

\textcolor{black}{The scattering parameters were measured using a two-port Vector Network Analyzer (VNA) calibrated in the frequency range from 100 MHz to 5 GHz, in which the transmitter was connected to port 1 and the receiver to port 2. Based on these measurements, the WPT efficiency $\eta$ can be calculated as:
\begin{equation}\label{eq7a}
  \eta = \frac{\left| S_{21}\right|^2}{1-\left| S_{11}\right|^2}
\end{equation}
in which the forward transmission coefficient $S_{21}$ is decorrelated from the input reflection coefficient $S_{11}$. These measurements were carried out for distances $d$ ranging from 50 mm to 200 mm.
}

\subsection{Results and Discussion}

\begin{figure}[htbp!]
\centering  
\includegraphics[scale = 0.96]{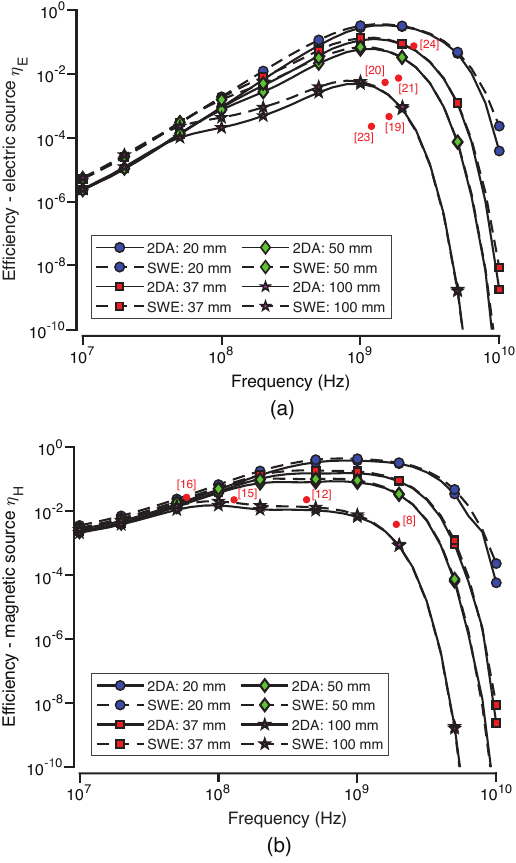}
\caption{\textcolor{black}{WPT efficiency for (a) an electric $\eta_E$ and (b) magnetic $\eta_H$ receiver as a function of the frequency and the phantom radius $R_p$ (asymptotically equivalent to the implantation depth) evaluated through the 2D-Axisymmetric (2DA) and the Spherical Wave Expansion (SWE) models. The red dots indicate the efficiency of physically realized WPT systems proposed in the literature and detailed in Table \ref{tab0}.}}
\label{fig2}
\end{figure}

\textcolor{black}{The WPT efficiencies for the electric $\eta_E$ and magnetic $\eta_H$ sources were evaluated from the electromagnetic fields calculated through the 2D-Axisymmetric model (2DA) and the Spherical Wave Expansion (SWE) model. In Fig. \ref{fig2}, the efficiency for both sources is shown as a function of the frequency for different phantom radius $R_p$, which is asymptotically equivalent to the variation in the implantation depth. Besides, the chosen range comprises most of the applications for wireless powered implants, from subcutaneous to deeply implanted devices. In addition, the measured efficiencies defined as $\eta = |S_{21}|^2$ from different WPT approaches in the literature is indicated in Fig. \ref{fig2} and they are detailed in terms of source type, operating frequency, and implantation depth in Table \ref{tab0}.} 

\begin{table}[htbp!]
\centering
\caption{\textcolor{black}{Comparison Between Different Implantable WPT Systems Proposed in the Literature}}
\label{tab0}
\begin{tabular}{c c c c c}
\hline\hline
\textbf{Reference} & \textbf{Source Type} & \textbf{Frequency} & \textbf{\begin{tabular}[c]{@{}c@{}}Implantation\\ depth\end{tabular}} & \textbf{Efficiency} \\ \hline\hline
\cite{8}  & Magnetic & 1.9 GHz   & 20 mm & 0.35\%  \\ 
\cite{49} & Magnetic & 433 MHZ   & 20 mm & 2.03\%  \\ 
\cite{47} & Magnetic & 131 MHz   & 18 mm & 2.01\% \\ 
\cite{48} & Magnetic & 60 MHz    & 16 mm & 2.40\%    \\ 
\cite{16} & Electric & 1.6 GHz   & 55 mm & 0.04\%  \\ 
\cite{17} & Electric & 1.5 GHz   & 55 mm & 0.45\%  \\ 
\cite{18} & Electric & 1.87 GHz  & 60 mm & 0.61\%  \\ 
\cite{20} & Electric & 1.2 GHz   & 58 mm & 0.02\%  \\ 
\cite{21} & Electric & 2.4 GHz   & 10 mm & 8\%     \\ \hline\hline
\end{tabular}
\end{table}

First of all, the results obtained from the two approaches previously described are compared. As it can be seen in Fig. \ref{fig2}, the efficiency evaluated with the 2D-Axisymmetric model agrees with the one calculated with Spherical Wave Expansion, especially at the vicinity of the optimum frequency where the error between them is 0.64\% for the electric receiver and 2.07\% for the magnetic one. However, for lower frequencies, an offset between them is noticeable mainly due to the difference in the source formulation. \textcolor{black}{Apart from that, the maximum efficiencies theoretically calculated with the proposed model corroborate with those experimentally achieved in previous works.}

Regarding the efficiency behavior over the frequency, for near-field operations at lower frequencies, the magnetic source outperforms the electric one for all evaluated implantation depths. Table \ref{tab1b} shows that once the optimum frequency for the magnetic receiver is below the far-field region, roughly starting at 1 GHz. In addition, at the near-field region, $\eta_E$ sharply increases until reaching its maximum, whereas $\eta_H$ initially increases then reaches a plateau at the maximum efficiency. Conversely, both receivers present similar decaying behavior in the far-field region, and $\eta_E$ asymptotically approaches $\eta_H$ at frequencies above the optimum one for the electric receiver.

\begin{table}[htbp!]
\centering
\caption{Maximum Efficiency and Optimum Frequency for the 2d-Axisymmetric Model as a Function of the Phantom's Radius (Asymptotic to the Implantation Depth)}
\label{tab1b}
\begin{tabular}{ccc}
\hline\hline
\textbf{$R_p$ (mm)} & \textbf{Electric Receiver} & \textbf{Magnetic Receiver} \\ \hline\hline
20               & 35.44\% at 1.26 GHz        & 44.92\% at 794.3 MHz       \\
37               & 13.43\% at 1.26 GHz        & 19.01\% at 398.1 MHz       \\
50               & 6.83\% at 1 GHz            & 10.44\% at 631 MHz         \\
100              & 0.49\% at 1 GHz            & 2.04\% at 100 MHz         \\ \hline\hline
\end{tabular}
\end{table}

As the reflection losses can be disregarded, this overall behavior for both receivers can be explained by the fact that in low frequencies, the near-field losses are higher, whereas, in the far-field region, the attenuation losses become more significant \cite{27}. Thus, the optimum frequency corresponds to the trade-off point between both loss mechanisms. Moreover, once the near-field losses affect more electric than magnetic sources, the efficiency at low frequencies is significantly lower for this receiver.

{\color{black} The analysis carried out in this paper considers fixed dimensions for the receiver. However, a further study on the peak efficiency and optimum frequency dependence on the receiver dimensions is presented in \cite{23}. It has shown that the receiver length $L$ is the parameter that impacts the efficiency the most; for instance, the larger the electric source, the higher the efficiency will be. On the other hand, for a magnetic source, the peak efficiency is reached within the range $1.5Rc < L < 3Rc$. In addition, the fundamental limit on the electrically small antennas efficiency states that a reduction in the receiver size will also lead to a reduction in its radiation efficiency. Therefore, the receiver size considered in the previous analysis not only is in the range that leads to maximum efficiency but also corresponds to the size of practical implementation of implantable and ingestible antennas \cite{37}.}

\textcolor{black}{The experimental results for the maximum efficiency and the optimum frequency considering an electric (dipole antenna) and a magnetic (loop antenna) transmitter are presented in Table~\ref{tab1c} for distances $d$ between the transmitter and receiver from 50 mm to 200 mm. The measurements agree with the theoretical conclusions obtained with the proposed model and the SWE approach, showing that as the distance between the transmitter and receiver (implantation depth) increases, the optimal frequency decreases for both sources. However, while the electric transmitter leads to an efficiency peak around 1 GHz which corresponds to the mid-field region, the magnetic transmitter achieves maximum efficiency for near-field operation. Moreover, the measurements were carried out in a solution with 1g of sodium chloride (NaCl) per liter. Although there are no significant changes on the optimum frequency, the increase in the medium conductivity leads to a reduction on the efficiency.  The difference in the efficiency values is due to the fact that the theoretical ones represent the maximum achievable efficiency taking into account only the intrinsic loss of the channel. In contrast, other losses, such as the scattering phenomena at the interface between the phantom and external media affect the experimental results (note that the antenna reflection losses are deembedded from the results; see Section~\ref{sec_ExpVal}).}


\begin{table}[htbp!]
\centering
\caption{\textcolor{black}{Measured Maximum Efficiency and Optimum Frequency as a Function of the Distance $d$ Between Transmitter and Receiver}}
\label{tab1c}
\begin{tabular}{cccc}
\hline\hline
\multirow{2}{*}{$d$ \textbf{(mm)}} &
  \multicolumn{3}{c}{\textbf{Electric Transmitter (Dipole)}} \\ \cline{2-4} 
 &
  \textbf{Optimum Frequency} &
  \textbf{\begin{tabular}[c]{@{}c@{}}Max. Efficiency\\ (Water)\end{tabular}} &
  \textbf{\begin{tabular}[c]{@{}c@{}}Max. Efficiency\\ (1 g/l NaCl)\end{tabular}} \\ \hline\hline
50 &
  1.15 GHz &
  $9.96 \times 10^{-4}$ &
  $3.51 \times 10^{-4}$ \\
100 &
  1 GHz &
  $2.02 \times 10^{-4}$ &
  $1.03 \times 10^{-4}$ \\
150 &
  953.6 MHz &
  $7.48 \times 10^{-5}$ &
  $1.75 \times 10^{-5}$ \\
200 &
  879.3 MHz &
  $3.69 \times 10^{-5}$ &
  $5.62 \times 10^{-6}$ \\ \hline\hline
\multirow{2}{*}{$d$ \textbf{(mm)}} &
  \multicolumn{3}{c}{\textbf{Magnetic Transmitter (Loop)}} \\ \cline{2-4} 
 &
  \textbf{Optimum Frequency} &
  \textbf{\begin{tabular}[c]{@{}c@{}}Max. Efficiency\\ (Water)\end{tabular}} &
  \textbf{\begin{tabular}[c]{@{}c@{}}Max. Efficiency\\ (1 g/l NaCl)\end{tabular}} \\ \hline\hline
50 &
  867 MHz &
  $1.53 \times 10^{-3}$ &
  $9.72 \times 10^{-5}$ \\
100 &
  743.2 MHz &
  $7.40 \times 10^{-4}$ &
  $3.46 \times 10^{-5}$ \\
150 &
  706 MHz &
  $2.49 \times 10^{-4}$ &
  $1.42 \times 10^{-5}$ \\
200 &
  693.7 MHz &
  $2.02 \times 10^{-4}$ &
  $7.52 \times 10^{-6}$ \\ \hline\hline
\end{tabular}
\end{table}

\section{Efficiency Assessment Using the Anatomical Pectoral Model}\label{secIII}

\textcolor{black}{To analyze the wireless power transfer efficiency in a more realistic scenario, we implemented a 2D heterogeneous model of the human pectoral region presented in Fig. \ref{fig1}(c) with nine dispersive media with complex permittivity equivalent to those in the different human tissues \cite{39}}. Based on the wirelessly powered pacemaker proposed in \cite{16}, the receiver is represented by a circular region with a radius of 1 mm inside the heart with implantation depth $d$ equal to 37 mm from the closest skin interface. 

Similarly, as described for the spherical model in Section \ref{secII}, the electric permittivity of the region external to the body is the same as a lossless skin ${\varepsilon}_{skin}(f)$, whereas the also lossless medium inside the pacemaker ${\varepsilon}_{muscle}(f)$ is perfectly matched to the wave impedance of the myocardium. In this way, the reflection losses can be disregarded.

For the analysis of this problem as a 2D model, it is assumed that the implantable receiver and the on-body transmitter are perfectly aligned, and therefore the $z$-axis invariance is valid, i.e., $\textbf{E}(x,y,z)=\textbf{E}(x,y)e^{ik_zz}$, where $k_z$ is the out-of-plane propagation constant. Moreover, due to the high tissue dispersivity, the radiated power is attenuated at short distances from the transmitter. Therefore, the computational cost can be reduced by analyzing only the cross-section of the body region in which the receiver is implanted.

The same approach described in Section \ref{secII} was applied to the 2D pectoral model to calculate the maximum WPT efficiency in a realistic scenario and compare these results with those obtained with the simplified spherical model. The reciprocity theorem is again invoked for calculation purposes, and the radiation source is considered inside the heart. This premise is still valid once the different media are linear and there is no polarization mismatch. Consequently, the obtained efficiency calculated with (\ref{eq3}) is valid for the original problem of powering the implantable device.

\begin{figure}[htbp!]
\centering  
\includegraphics[scale = 1]{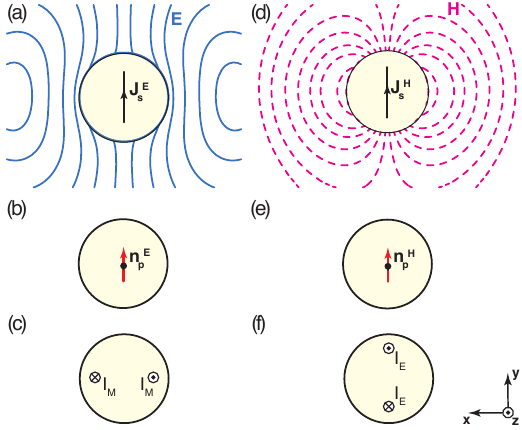}
\caption{Different receiver configurations are used for the realistic model analysis. An electric receiver can be defined through (a) a surface electric current distribution $\mathbf{J}_s^E$, (b) an electric point dipole $\mathbf{n}_p^E$, and (c) an out-of-plane magnetic current $I_M$. Alternatively, a magnetic receiver can be modeled as (d) a surface magnetic current distribution $\mathbf{J}_s^H$, (e) a magnetic point dipole $\mathbf{n}_p^H$, and (f) an out-of-plane line current $I_E$. The contour lines depict the electric $\mathbf{E}$ and magnetic $\mathbf{H}$ field distribution in arbitrary units.}
\label{fig3}
\end{figure}

However, differently from the 2D-axisymmetrical case, different receiver configurations and orientations can be used in the realistic planar model, as shown in Fig. \ref{fig3}.  The first possibility is to assign a Surface Current Distribution (SCD) over the line of length $L$ (in the 2D simulation) for the electric $\textbf{J}_s^E(x,y,z)$ and magnetic $\textbf{J}_s^H(x,y,z)$ receivers according to:

\begin{subequations}
\begin{equation}\label{eq8a}
  \textbf{J}_s^E=\begin{cases}
\cos\left ( \frac{\pi y}{L} \right ) \; \textup{A/m} \; \mathbf{\hat{y}}, \text{ if vertically oriented}\\ 
\cos\left ( \frac{\pi x}{L} \right ) \; \textup{A/m} \; \mathbf{\hat{x}}, \text{ if horizontally oriented} 
\end{cases}
\end{equation}    
\begin{equation}\label{eq8b}
\textbf{J}_s^H=\begin{cases}
1 \; \textup{V/m} \; \mathbf{\hat{y}},  \text{ if vertically oriented}\\ 
1 \; \textup{V/m} \; \mathbf{\hat{x}},  \text{ if horizontally oriented} 
\end{cases}
\end{equation}
\end{subequations}

Henceforward, vertically oriented means that the receiver is oriented in the $y$-axis direction ($\mathbf{\hat{y}}$) whereas horizontally in the $x$-axis direction ($\mathbf{\hat{x}}$), according to the coordinate system depicted in Fig. \ref{fig3}. 

Subsequently, the next alternative is to define the receiver as an Electric Point Dipole (EPD) or Magnetic Point Dipole (MPD). In each one, the electric $\textbf{n}_p^E$ or magnetic $\textbf{n}_p^H$ dipole moment vector is assigned as follows:

\begin{subequations}
\begin{equation}\label{eq9a}
\mathbf{n}_p^E=\begin{cases}
1 \; \textup{A\( \cdot \)m} \; \mathbf{\hat{y}},  \text{ if vertically oriented}\\ 
1 \; \textup{A\( \cdot \)m} \; \mathbf{\hat{x}},  \text{ if horizontally oriented} 
\end{cases}
\end{equation}    
\begin{equation}\label{eq9b}
\mathbf{n}_p^H=\begin{cases}
1 \; \textup{A\( \cdot \)m}^2 \; \mathbf{\hat{y}},  \text{ if vertically oriented}\\ 
1 \; \textup{A\( \cdot \)m}^2 \; \mathbf{\hat{x}},  \text{ if horizontally oriented} 
\end{cases}
\end{equation}
\end{subequations}

Finally, it is also possible to define an out-of-plane Magnetic Current (MC) for modeling the electric receiver or an out-of-plane Line Current (LC) for the magnetic one. Respectively for each case, the electric or magnetic fields are orthogonal to the current distribution direction and the position vector. Therefore, the out-of-plane magnetic $I_M$ = 1 V and electric $I_E$ = 1 A currents must be assigned following the direction and orientation shown in Fig. \ref{fig3}(c) and Fig. \ref{fig3}(f), respectively.

\begin{figure}[htbp!]
\centering  
\includegraphics[scale = 1]{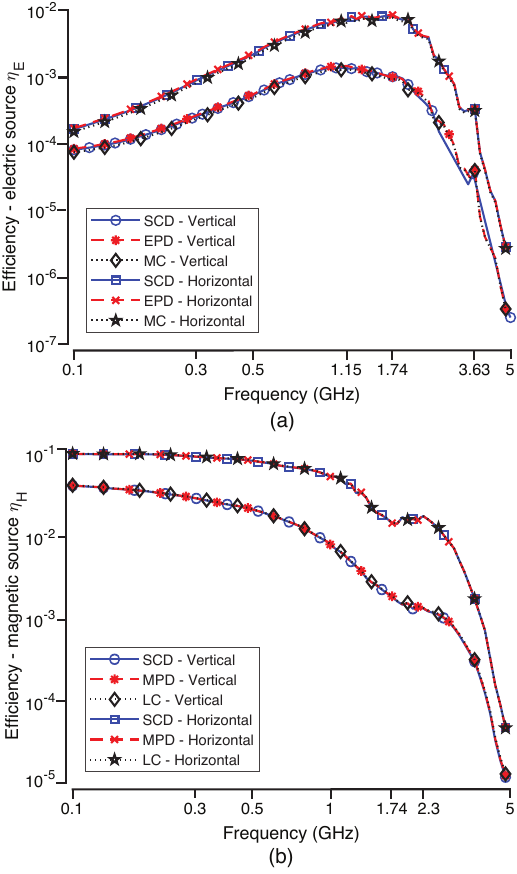}
\caption{Efficiency was evaluated with the realistic pectoral model for an (a) electric $\eta_E$ and (b) magnetic $\eta_H$ receiver implanted in the heart. Different receiver orientations (vertical and horizontal) and configurations \{Surface Current Distribution (SCD), Electric (EPD) and Magnetic Point Dipole (MPD), out-of-plane Magnetic Current (MC) and Line Current (LC)\} are compared.}
\label{fig4}
\end{figure}

By taking into account the three source types and the vertical and horizontal orientations, six configurations for each electric and magnetic receiver are possible. Finally, the WPT efficiency was evaluated for all the described scenarios, and the results are presented in Fig. \ref{fig4}. It is possible to see that the efficiencies evaluated with different sources agree between them, which demonstrates the flexibility of the proposed model for dealing with different source configurations. 

The source orientation is the factor that mainly affects the radiation efficiency. In the best-case scenario, the sources are horizontally-oriented, whereas the vertical orientation corresponds to the worst position for the receiver. Since the receiver is vertically oriented, the radiation beam propagates towards the direction parallel to the chest boundary, which is the closest to the receiver and, therefore, the region that contributes the most for the integral in (\ref{eq2b}). This behavior was also pointed out in \cite{22} but, from the results shown in Fig. \ref{fig4}, it is possible to evaluate the difference between best and worst-case scenarios quantitatively. Namely, for the best positioning, the electric source presents an optimum frequency 590 MHz higher with a maximum $\eta_E$ approximately five times the one verified with a vertically oriented source. A similar offset in efficiency magnitude can be verified between a horizontal and a vertical magnetic source. For lower frequencies, the best positioning leads to an efficiency around 2.5 times higher; however, the maximum difference between them surpasses 13 times in the far-field region.

It is also noticeable that the radiation efficiency behavior for an electric or magnetic receiver obtained with the realistic pectoral model agrees with those obtained with the simplified spherical model.  Namely, for the horizontally oriented electric sources, $\eta_E$ increases in the near-field range with a peak of 0.77\% at 1.74 GHz, close to the optimum frequency shown in Table \ref{tab1b} for the same implantation depth of 37 mm, then decreasing for higher frequencies. In contrast, $\eta_H$ reaches its maximum at lower frequencies, remaining almost constant around 10\% in the near-field, before decreasing for frequencies above 1 GHz. The sharp reduction in radiation efficiency by a magnetic source in the far-field region is explained by the fact that this kind of source achieves its optimal performance as an electrically small antenna \cite{42},\cite{43}, i.e., when $L < \lambda/10$ which is satisfied only in the near-field region.

As expected, there is a significant reduction in the maximum efficiencies obtained with the realistic model compared with the simplified spherical model. First of all, the latter considers a homogeneous medium equivalent to the human muscle tissue, whereas in the pectoral model, several organs with different electromagnetic properties are taken into account. Two predominant loss sources are responsible for this efficiency deterioration: the increased attenuation, mainly in the fat and muscle layers where the power is dissipated before reaching the receiver in the heart; and the reflection losses due to the wave-impedance contrast between the media that compose the tissues. In this last case, with the multiple reflections, most of the reflected power is dissipated in the tissue reducing the amount of power that reaches the receiver encapsulation. Therefore, this loss is accounted on (\ref{eq1}) also through $P_d$. Moreover, in a realistic scenario, the shape of the organs may lead to resonance modes responsible for local peaks and valleys in efficiency for some frequencies, such as 3.63 GHz for the electric source and around 1.74 GHz for the magnetic source.

\section{Efficiency Assessment Using a Physically Realizable Model}\label{secIV}

\begin{figure}[htbp!]
\centering  
\includegraphics[scale = 0.65]{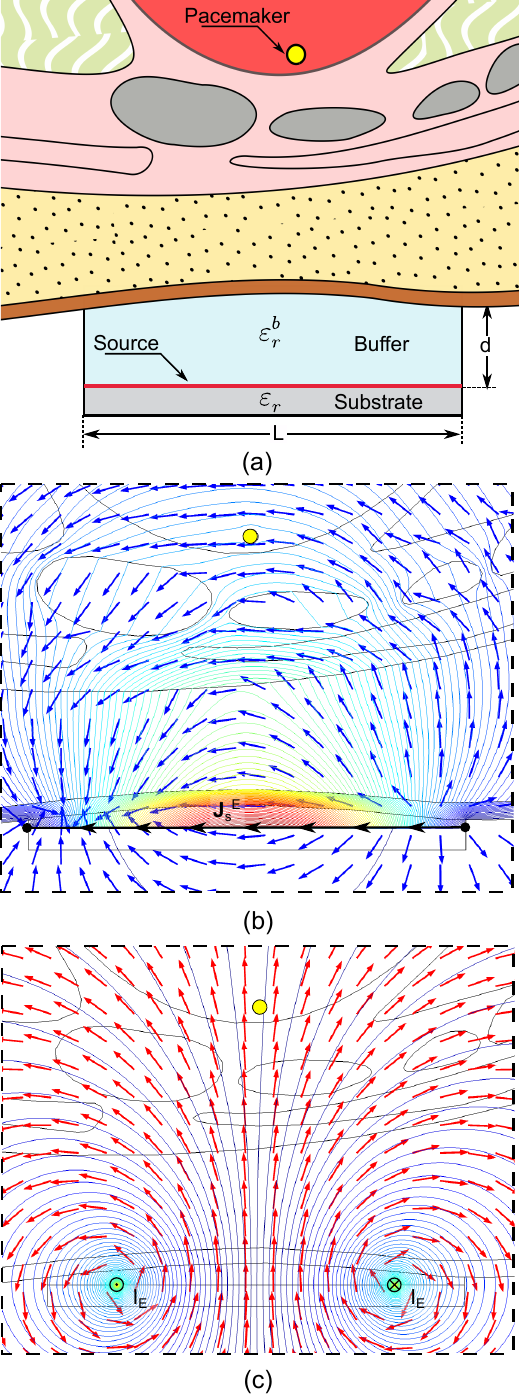}
\caption{Analysis setup for the parametric analysis of the efficiency in a wireless powered pacemaker application: (a) close-up view of the on-body transmitter with length $L$ composed of a radiation source positioned apart from the skin by a lossless superstrate (buffer) with thickness d and permittivity $\varepsilon_r^b$ and with a substrate matched to the medium external to the body. (b) Electric source modeled as a surface current density $\mathbf{J}_s^E$ with the electric field distribution indicated by the arrows and the magnetic field by the isolines. (c) Magnetic source modeled as an out-of-plane current $\mathbf{I}_E$ with the magnetic field distribution indicated by the arrows and the electric field by the isolines.}
\label{fig5}
\end{figure}

The case study of a wirelessly powered pacemaker is analyzed to establish the optimum parameters that lead to the maximum efficiency in a physically realizable application. In this case, the human pectoral model in Fig. \ref{fig1}(c) is used. However, the transmitter is now localized in the chest, as shown in Fig. \ref{fig5}(a). This transmitter comprises a substrate with the same dielectric properties as the medium external to the body and a lossless superstrate (or buffer) with thickness $d$ and relative electric permittivity $\varepsilon_r^b$.

The electric transmitter in Fig. \ref{fig5}(b) is defined by assigning a surface current distribution $\textbf{J}_s^E(x,y,z)$ given by:
\begin{equation}\label{eq10}
    \textbf{J}_s^E = \left [ \cos\left (\frac{\pi x}{L}  \right ),0,0 \right]
\end{equation}
in which $L$ is the transmitter length. Concurrently, the magnetic transmitter is modeled as a unitary out-of-plane electric current $I_E$ = 1 A on the edge of the source with opposite phases, as illustrated in Fig. \ref{fig5}(c). As seen from the field distribution for the electric and magnetic transmitters, they are, for instance, equivalent to a dipole and loop antenna, respectively.

\subsection{WPT Efficiency as Function of Transmitter Parameters}

In this configuration, the two main transmitter parameters that can be changed to achieve the maximum WPT efficiency are the buffer thickness $d$ and its permittivity $\varepsilon_r^b$. Therefore, a parametric analysis was carried out for both parameters within the ranges: $1 \leq d \leq 50$ mm and $1 \leq \varepsilon_r^b \leq 80$. Furthermore, the lower and upper bounds for $\varepsilon_r^b$ were chosen so that this range comprises the permittivities of all nine tissues in the frequency range from 100 MHz to 5 GHz. Consequently, the effective permittivity for this pectoral region is also in the chosen interval.

The operating frequency must be chosen and fixed to carry out this parametric study. Based on the results shown in Fig. \ref{fig4}, the maximum theoretical WPT efficiency for a deep-implanted pacemaker charged by an electric source transmitter is achieved at the frequency of 1.74 GHz. Considering the frequency bands allocated for Industrial, Scientific, and Medical (ISM) usage, the closest band to the optimal frequency is centered at 2.45 GHz. Henceforward, this is the frequency chosen for all following analysis, and the transmitter length $L$ was set as $\lambda/2$ at this frequency in free space. 

In order to evaluate the WPT system for this configuration, (\ref{eq3}) can still be applied with the results from integrals in (\ref{eq2a}-\ref{eq2c}); however, the integration domains need to be reformulated. In this case, $\Omega_s$ indicates that the integral is performed on the boundary of the pacemaker, whereas $\Omega_p$ represents the transmitter's boundary.

\begin{figure}[htbp!]
\centering  
\includegraphics[scale = 0.95]{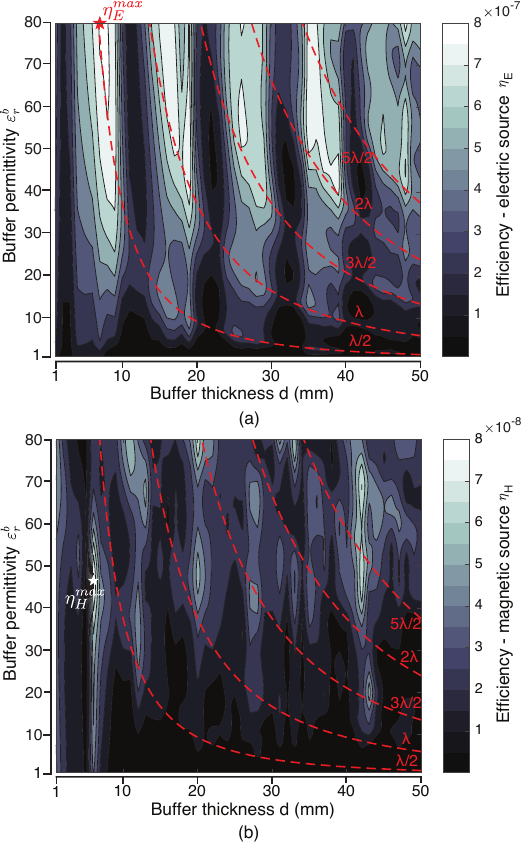}
\caption{Contour plots for the efficiency as a function of the buffer thickness $d$ and permittivity $\varepsilon_r^b$ for an (a) electric source $\eta_E$ and a (b) magnetic source $\eta_H$. The star indicated the global maximum position, and the dotted lines indicate the position corresponding to $\lambda/2$ its integer multiples as a function of $\varepsilon_r^b$.}
\label{fig6}
\end{figure}

The contour levels for the efficiency as a function of the buffer thickness and permittivity are shown in Fig. \ref{fig6}(a) for electric and Fig. \ref{fig6}(b) for magnetic sources. As it can be seen, the buffer parameters that lead to the maximum efficiency are $d =$ 7 mm and $\varepsilon_r^b =$ 80. The fact that the maximum efficiency is obtained with the highest permittivity can be explained once the wave-impedance mismatching at the buffer-skin transition is practically eliminated. In addition, taking into account this permittivity, the maximum occurs exactly at $\lambda/2$. As proved in \cite{44} by considering the superstrate as a transmission line, it corresponds to the resonance condition, once the impedance seen in the transmitter-skin transition is equal to the impedance seen in the source, which leads to a high voltage at the source position, thus increasing the radiation efficiency. Moreover, it can be seen that as the buffer's thickness increases, so decreases the optimum $\varepsilon_r^b$ in such a manner that a $\lambda/2$ periodicity is obeyed, as it also happens with a lossless transmission line.

On the other side, the maximum efficiency obtained with a magnetic source is approximately an order of magnitude lower when compared with the electric one. It means that, at the frequency of 2.45 GHz, the magnetic source is sub-optimal, as it was also predicted in the graph of Fig. \ref{fig4}(b), once it does not satisfy the electrically small antenna condition at this frequency. However, the maximum efficiency is achieved with $d = 6$~mm and $\varepsilon_r^b= 46.5$. It is important to mention that, in this case, this maximum occurs in a significantly narrow distance range, which in practice would require precise positioning of the source. 

\subsection{Maximum Efficiency Assessment due to Stochastic Antenna Misalignment}

In order to classify and quantify the losses, the optimal scenario in which the transmitter is composed of an electric source and a buffer with $d =$ 7 mm and $\varepsilon_r^b =$ 80 is considered. Three main loss mechanisms can be identified: reflection losses due to the wave impedance contrast between the on-body transmitter and the body tissue, antenna near-field losses, and the attenuation losses in the tissues. First two losses can be mitigated by adequately designing the WPT system, whereas the latter cause is unavoidable. For instance, approximately 7\% of the transmitted power is lost due to reflection in this optimum configuration, whereas 93\% is due to attenuation and near-field losses combined. However, the optimal buffer parameters justify the low reflection losses and reduce the near-field region and its associated losses.

\begin{figure}[htbp!]
\centering  
\includegraphics[scale = 1]{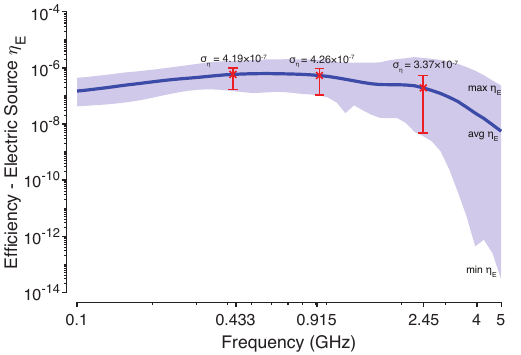}
\caption{\textcolor{black}{WPT efficiency considering an optimal electric transmitter. The continuous blue line is the efficiency averaged (avg $\eta_E$) over the 200 pacemaker positions. Besides, the blue area represents the maximum (max $\eta_E$) and minimum (min $\eta_E$) efficiency values. Finally, the red bars show the standard deviation on the efficiency $\sigma_\eta$ at the ISM frequencies of 433 MHz, 915 MHz, and 2.45 GHz.}}
\label{fig6b}
\end{figure}

\textcolor{black}{Apart from that, the maximum efficiency levels are significantly affected by fluctuations in the implantation depth caused by the intrinsic movements of the human body. For instance, the heart movements lead to fluctuations in the implant position, resulting in misalignment between the transmitter and receiver. Therefore, in order to evaluate these efficiency variations, the full-wave simulation was performed at 200 random implant locations uniformly distributed within the heart. After that, the WPT efficiency for each pacemaker position is calculated, considering the optimal electric transmitter. The obtained results for maximum, minimum, and average efficiency are shown in Fig. \ref{fig6b}. In addition, the standard deviations on the efficiency values $\sigma_\eta$ are shown for the ISM frequencies of 433 MHz, 915 MHz, and 2.45 GHz.}

\textcolor{black}{The results in Fig. \ref{fig6b} show that the difference between the minimum and maximum efficiency values increases with the frequency; however, the average efficiency and standard deviation remain within a close range for the frequency bands commonly used in biomedical applications. This variation in the optimum frequency is explained by the fact that the optimum frequency and the maximum efficiency decrease as the implantation depth increases, as previously described in Fig. \ref{fig2}.}

Finally, the fact that the efficiency obtained in this case is significantly lower when compared with the optimum efficiency presented in Fig. \ref{fig4} shall be addressed. It can be explained by the fact that the transmitter is considered to be a conformal structure covering the entire body surface in the ideal scenario described in Section \ref{secIII}. In contrast, the current analysis takes into account a mono-antenna transmitter, leading to a much smaller integration domain $\Omega_p$ thus reducing the received power. In practical terms, it means that the single antenna mid-field WPT solutions currently in the literature are sub-optimal by many orders of magnitude. Therefore, to physically achieve the maximum efficiency theoretically obtained, further developments in mid-field WPT are required based on conformal antenna arrays such as in \cite{45} and techniques for electromagnetic focusing in lossy media that have already been investigated for other biomedical applications \cite{46}.

\subsection{Electromagnetic Exposure Assessment}

\begin{figure}[htpb!]
\centering  
\includegraphics[scale = 0.95]{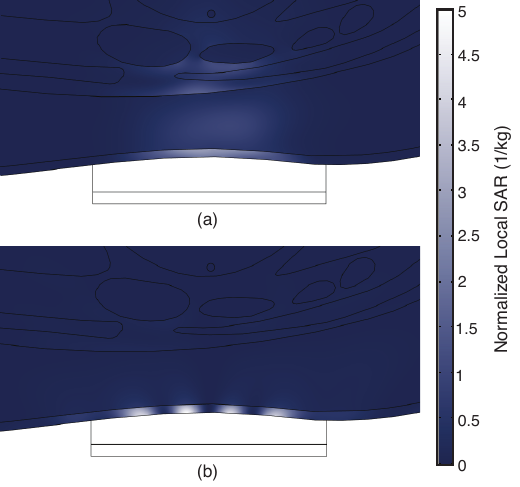}
\caption{Surface plot of the local specific absorption rate (point SAR) in the region between the on-body transmitter and the deep-implanted receiver normalized by the transmitted power considering an optimal (a) electric and (b) magnetic source.}
\label{fig7}
\end{figure}

\textcolor{black}{Another major concern of in-body WPT applications is the electromagnetic exposure levels.} The surface profiles for the local specific absorption rate (SAR) normalized by the transmitted power are shown in Fig. \ref{fig7} for the optimal electric and magnetic sources. First of all, the SAR profiles follow the field distribution shown in Fig. \ref{fig5}(b) and Fig. \ref{fig5}(c), being higher in the center for the electric source, whereas it reaches its maximum around the source edges for the magnetic. In addition, in both cases, the peak SAR is located in the skin layer, and no significant SAR values are observed in the vital organs. However, by comparing both SAR levels, it is noticed that the magnetic source leads to a larger SAR, almost twice the value for the electric source, even with much lower efficiency levels. Since the SAR is related to the attenuation in the tissues that cannot be physically mitigated, it is possible to conclude that the efficiency is maximized by properly choosing the transmitter parameters, and the electromagnetic exposure is minimized.

\section{Guidelines for Designing Optimal Implantable WPT Systems}

\textcolor{black}{The analysis and results described in the previous sections lead to useful insights that can be applied as guidelines to design implantable WPT systems that are able to achieve efficiency close to the expected optimal values. For instance, given the implantation depth of the receiver, the following points can be considered:}

\subsection{WPT Technique and Operating Frequency}
\textcolor{black}{For subcutaneous implants, near-field techniques such as inductive and capacitive coupling can achieve higher efficiency levels. However, as the implantation depth increases, the maximum efficiency is reached at frequencies between 900 MHz and 3 GHz. Therefore, mid-field and far-field techniques are more indicated for powering deep-implanted bioelectronics. Moreover, given the implantation depth and the operating frequency, the maximum efficiency values are given in Fig.~\ref{fig2}.}

\subsection{Antenna Type and Orientation}
At frequencies below 300 MHz, magnetic loop antennas are able to reach significantly higher efficiencies than electric-type antennas. On the other side, the high permittivity of the human tissues electrically increases the effective aperture size of dipole or patch antennas, making them more suitable for implantable WPT systems operating at frequencies higher than the sub-GHz band. The maximum efficiency values for both electric and magnetic antennas as a function of the frequency are presented in Fig. \ref{fig2} whereas the effect of the source orientation is shown in Fig. \ref{fig4}.

\subsection{On-body Transmitter Design}
\textcolor{black}{For single on-body transmitter applications, higher WPT efficiencies can be achieved by designing a buffer structure that separates the transmitter antenna and the skin, as shown in Fig. \ref{fig6}. In this case, the relative permittivity and the thickness of this superstrate must be chosen in order to reduce the wave-impedance mismatch at the transmitter-skin interface~\cite{rice_high-contrast_2022}. For instance, for an electric source, the maximum WPT efficiency is achieved for the buffer thickness equal to $\lambda/2$. Apart from that, the standard deviation on the efficiency due to stochastic misalignments between the on-body transmitter and the in-body receiver caused by body movements is shown in Fig. \ref{fig6b} for the main ISM frequencies used in biomedical applications. Moreover, the local SAR patterns for an electric and magnetic on-body transmitter can be seen in Fig. \ref{fig7}. Even though the single-antenna configuration is constructively simpler, it exhibits suboptimal performance for implantable WPT systems. Therefore, in order to achieve efficiency closer to the levels given in Fig. \ref{fig2} and lower exposure levels, conformal transmitter arrays should be employed in addition to applying techniques for electromagnetic focusing in lossy media.}

\subsection{In-body Receiver Design}
\textcolor{black}{Although the constraints on the shape and dimensions of in-body antennas reduce the degrees of freedom in their design, some points can be observed in order to enhance the receiver performance. First, an appropriate type of the antenna (electric or magnetic) has to be chosen depending on the receiver size and its operating frequency~\cite{23}. A high-permittivity encapsulation material can be used to increase the efficiency of an electric-type receiver~\cite{nikolayev_robust_2017}. The dielectric loading electrically increases the aperture size, reduces the near-field losses, and an efficiency close to the fundamental bounds can be achieved \cite{23}. Other alternatives to enhance the efficiency are filling the receiver encapsulation with a low-loss dielectric material and using the thickest possible substrate \cite{37}. Finally, reconfigurable receiver antennas can be implemented to achieve near-optimal radiation performance across a broad range of conditions \cite{30_1}, \cite{bao_conformal_2018}.}

\section{Conclusion}\label{secV}
In order to wirelessly power implantable bioelectronic devices, several techniques can be employed depending on the application characteristics. However, the combination of different physical mechanisms imposes a limit on maximum achievable wireless powering efficiency. Among them, the principal mechanism is the electromagnetic attenuation in the dispersive tissues, the reflection losses, and the fundamental efficiency limitation of electrically small antennas. In addition to it, there are system-level losses and resonant modes in the body cavities that also contribute to efficiency deterioration. Therefore, this work focused on evaluating the wireless powering efficiency of deep-body implanted devices and studying how the system’s parameters could be tuned to reach the highest efficiency possible in a practical application.

Towards this goal, a mathematical formulation for this problem was proposed based on reciprocity and T-symmetry principles as well as considering an on-body conformal source, perfectly matched to the human tissue. Such hypotheses lead to optimum energy focusing and, thus, to the highest possible efficiency. In order to validate this approach, a semi-analytical Spherical Wave Expansion analysis was carried out, and the results for a simplified spherical representation for the human body were compared.

This first analysis indicates that a magnetic receiver outperforms the electric one in the near-field frequency range, whereas, in the far-field, both receivers show similar performance. Therefore, the optimum frequency is in the mid-field region and depends on the source type and implantation depth, leading to the highest efficiency. 

Once the proposed approach was validated, it was applied to an anatomical model focused on a deep-implanted wireless powered pacemaker application \cite{16}. This analysis verified that the overall behavior for both sources agrees with the one predicted by the simplified spherical model, including the optimum operation in the mid-field range. Furthermore, this optimum frequency also agrees with the results presented in the literature obtained through different approaches \cite{22},\cite{23},\cite{25,26,27,28,29}. Even though there is a significant efficiency reduction in the realistic model results, the wave impedance contrast between the different tissues imposes reflection losses that were disregarded in the simplified model. Moreover, with this model, it was verified that the receiver orientation is another practical aspect that can reduce efficiency. Considering the source positioning, the maximum difference in efficiency between the best and worst-case scenarios is around five times for an electric receiver, whereas, for the magnetic one, this difference can be above 13 times. The effect of the source orientations is justified once in the best case, i.e., for a horizontally oriented receiver, the power flows towards the chest line, which is the closest point to the receiver, contributing more to the overall received power. It is important to mention that the flexibility in analyzing different source types is another contribution of the proposed approach compared to other methods of analysis.

Finally, a WPT system considering a deep-implanted pacemaker as the receiver and a multilayered (substrate-source-buffer) transmitter positioned in the chest was considered to analyze the impact of the transmitter’s parameters on the powering efficiency as well as to provide information on how these parameters could be tuned in order to achieve the maximum efficiency. This parametric study has shown that in the 2.4 GHz ISM band, the electric transmitter leads to efficiency over a degree of magnitude higher than a magnetic one. In addition, the maximum efficiency can be achieved with a lossless high-permittivity buffer, matched to the skin permittivity, and thickness equal to $\lambda/2$ in this medium. Furthermore, other local maxima can be obtained for thicker and lower permittivity superstrates, respecting a $\lambda/2$ periodicity.

To summarize, the approach proposed in this paper is able to predict the maximum achievable powering efficiency considering simplified or anatomical body models as well as to provide information on the parameters that can be optimized in order to achieve this maximum efficiency. In addition, even though the case study of a deep-implanted pacemaker was analyzed in this paper, the same approach could also be used to evaluate the powering efficiency for other deep-body implantable devices.

\balance

\bibliographystyle{IEEEtran}
\bibliography{bibliography.bib}

\end{document}